\documentclass[12pt]{article}
\usepackage{amsmath}
\usepackage{amsfonts}
\usepackage{mathptm}
\usepackage{epsfig}
\usepackage{pst-plot,epsf}
\newcommand{\beq}{\begin{equation}}
\newcommand{\eeq}{\end{equation}}
\newcommand{\bea}{\begin{eqnarray}}
\newcommand{\eea}{\end{eqnarray}}
\newcommand{\nn}{\nonumber}

\newcommand{\epm}{e^+e^-}

\newcommand{\ra}{\rightarrow}

\begin{document}
\thispagestyle{empty}
\begin{flushright}
May 2013\\
Revised:\\
July 2013\\
\vspace*{1.5cm}
\end{flushright}
\begin{center}
{\LARGE\bf {\tt carlomat,}\\[4mm]
version 2 of the program for automatic computation of lowest order cross 
sections}\\
\vspace*{2cm}
Karol Ko\l odziej\footnote{E-mail: karol.kolodziej@us.edu.pl}\\[1cm]
{\small\it
Institute of Physics, University of Silesia\\ 
ul. Uniwersytecka 4, PL-40007 Katowice, Poland}\\
\vspace*{3.5cm}
{\bf Abstract}\\
\end{center}
Version 2 of {\tt carlomat}, a program for automatic 
computation of the lowest order cross sections of multiparticle reactions, is
described. The substantial modifications with respect to 
version 1 of the program include: 
generation of a single phase space parametrization 
for the Feynman diagrams of the same topology,
an interface to parton density functions, improvement of the color matrix 
computation, the Cabibbo-Kobayashi-Maskawa mixing
in the quark sector, the effective models including
scalar electrodynamics, the Wtb interaction with operators 
of dimension up to 5 and a general top-higgs coupling. 
Moreover, some minor modifications have been made and several bugs in the 
program have been corrected. 
\vfill
\newpage
{\large \bf PROGRAM SUMMARY}\\[4mm]
{\it Program title:} {\tt carlomat, version 2.0}\\
{\it Catalogue identifier:}\\
{\it Program summary:}\\
{\it Program obtainable from:} CPC Program Library, Queen's University, Belfast,
N. Ireland\\
{\it Licensing provisions:} Standard CPC licence\\
{\it No. of lines in distributed program, including test data, etc.:}\\
{\it No. of bytes in distributed program, including test data, etc.:}\\
{\it Distribution format:} tar.gz\\
{\it Programming language:} {\tt Fortran 90/95}\\
{\it Computer:} all\\
{\it Operating system:} Linux\\
{\it Classification:}\\
{\it Nature of problem:} Leading order predictions for reactions of two
particle scattering into a final state with
up to 10 particles within the Standard Model and some effective models.\\
{\it Reasons for the new version:}
To adjust the program for description of hadron collisions,
facilitate computation of the colour matrix that is usually much
more involved for processes of the hadron--hadron collision than 
for processes of electron--positron annihilation,
shorten compilation time of the generated kinematical routines
and implement some extensions of the standard model in the program.\\
{\it Summary of revisions:}
A few substantial modifications are introduced with respect to
version 1.0 of the program. First, a single phase space parametrization
is generated for the Feynman diagrams of the same topology, taking into
account possible differences in mappings of peaks in the individual
diagrams, which speeds up a compilation time of the Monte Carlo program
for multiparticle reactions by a factor 4-5 with respect to the previous
version. Second, an interface to parton density functions is added that
allows to make predictions for hadronic collisions. Third, 
calculation of the color matrix is facilitated. Fourth,
the Cabibbo-Kobayashi-Maskawa mixing in the quark sector is implemented.
Fifth, the effective models including
scalar electrodynamics, the Wtb interaction with operators of dimension
up to 5 and a general top-higgs coupling are implemented. Moreover, some
minor modifications have been made and several bugs in the
program have been corrected.\\
{\it Method of solution:}
As in version 1 of the program, the matrix element in the helicity basis and 
multichannel Monte Carlo phase space integration routine are generated 
automatically for a user specified process. 
The color matrix is divided into smaller routines and written down 
as a stand alone program that is calculated prior to compilation and execution 
of the Monte Carlo program for computation of the cross section. 
The phase space integration routine is substantially shortened in order 
to speed up its compilation. The code generation part of 
the program is modified to incorporate the scalar electrodynamics and effective
Lagrangians of the top quark interactions with the $W$ and higgs bosons. 
Routines necessary for computing the helicity amplitudes of new couplings 
are added.\\
{\it Restrictions:}
Although a compilation time has been shortened in the current version, it 
still may be quite long for processes with 8 or more final state particles.
Another limitation is a size of the color matrix that, if too big, may
prevent compilation or result in a very long execution time of the
color computation program. This actually may happen already
for some QCD processes with 7 partons such as $gg\to 5g$, the
commutation time of the color matrix of which is about 200 hours.\\
{\it Running time:} Depends strongly on the selected process and, to less
extent, on the Fortran compiler used. 
The following amounts of time were needed  at different computation stages 
of the top quark pair production parton level process
$g g \;\to\; b u \bar{d} \;\bar{b} \mu^- \bar{\nu}_{\mu}$ 
to produce the appended test output files 
on a PC with the Pentium~4 3.0~GHz processor with Absoft 
(GNU, Intel) Fortran compilers: code generation 
takes 3.7~s (3.7~s, 2.4~s), compilation, computation and simplification of 
the color matrix takes about 1~s (1~s, 1~s), 
compilation of all the generated routines takes just a few seconds and
execution of the Monte Carlo program takes about 44~s (41~s, 23~s).

\vspace*{1.5cm}
{\large \bf LONG WRITE-UP}
\section{Introduction}
\begin{figure}[htb]
\centerline{
\epsfig{file=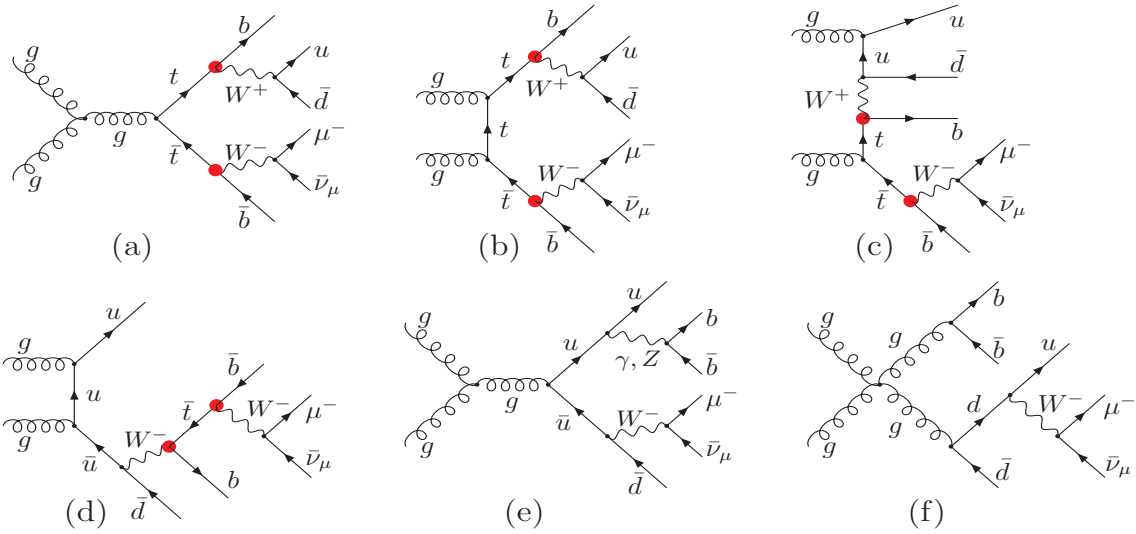,  width=150mm, height=70mm}}
\caption{Examples of the leading order Feynman diagrams of process
(\ref{ggbbudmn}). Blobs indicate the $Wtb$ coupling.}
\label{diags_gg}
\end{figure}

\begin{figure}[htb]
\centerline{
\epsfig{file=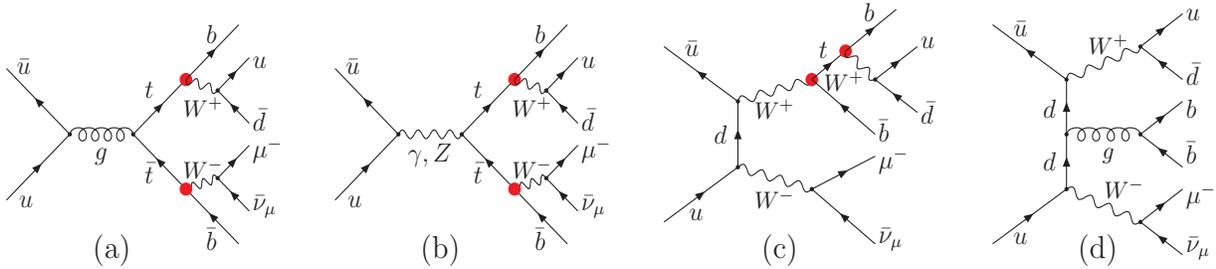,  width=160mm, height=35mm}}
\caption{Examples of the leading order Feynman diagrams of process
(\ref{uubbudmn}). Blobs indicate the $Wtb$ coupling.}
\label{diags_uu}
\end{figure}

In the era of the LHC,
production of a few heavy particles such as the
electroweak (EW) gauge bosons, top quark or recently discovered scalar candidate
for the higgs boson, at a time have become commonplace.
The heavy particles live so shortly that they should be actually regarded
as the intermediate states of reactions with multiple light particles
in the final state. 
As an example consider
\bea
\label{ggbbudmn}
gg &\ra& b u\bar{d}\;\bar{b} \mu^-\bar{\nu}_{\mu},\\
\label{uubbudmn}
u\bar u &\ra& b u\bar{d}\;\bar{b} \mu^-\bar{\nu}_{\mu}
\eea
which are underlying partonic processes of the top quark 
pair production in the proton-proton collisions at the LHC, $pp\to t\bar t$.
The final state of (\ref{ggbbudmn}) and (\ref{uubbudmn})
corresponds to each of the top quarks decaying into
a $b$-quark and a $W$-boson and, subsequently one of the $W$ bosons decaying 
leptonically and the other hadronically, as depicted in Figs.~\ref{diags_gg}a,
\ref{diags_gg}b, \ref{diags_uu}a and \ref{diags_uu}b. However, both reactions
receive contributions from many other Feynman diagrams which do not represent
the signal of the top quark pair production. Some examples of such diagrams
are depicted in Figs.~\ref{diags_gg}c--\ref{diags_gg}f for reaction
(\ref{ggbbudmn}) and in Figs.~\ref{diags_uu}c--\ref{diags_uu}d for reaction
(\ref{uubbudmn}).
The entire number of the leading order Feynman diagrams of reactions 
(\ref{ggbbudmn}) and (\ref{uubbudmn}) in the unitary gauge of the standard
model (SM), with the neglect of the higgs boson coupling to fermions
lighter than the $b$-quark and of the Cabibbo-Kobayashi-Maskawa (CKM) mixing
amounts to 421 and 718, respectively. 

\begin{figure}[htb]
\centerline{
\epsfig{file=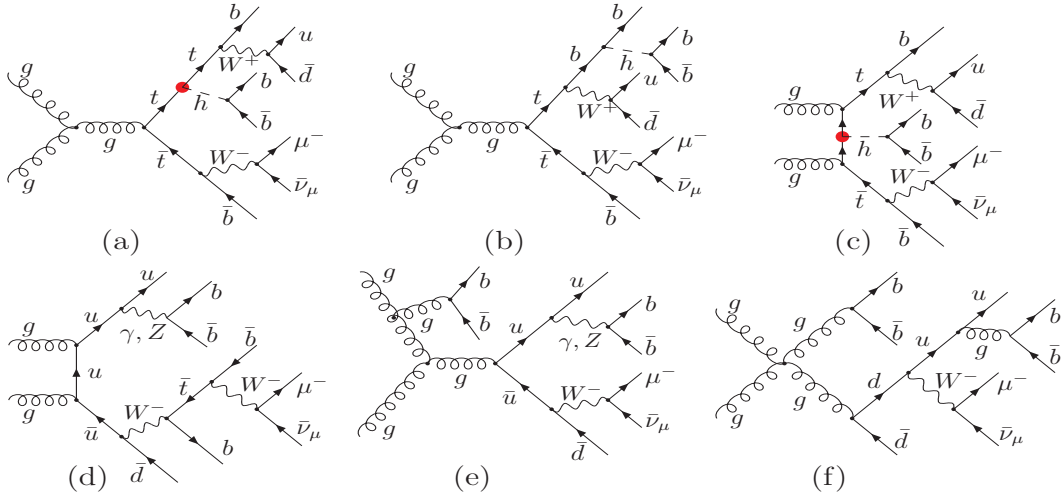,  width=140mm, height=65mm}}
\caption{Examples of the lowest order Feynman diagrams of reaction 
(\ref{ggbbbudbmn}): (a), (b) and (c) are the signal diagrams of $t\bar t h$ 
production, (d), (e) and (f) are the $t\bar t h$ background contributions. 
Blobs indicate the higgs--top coupling.}
\label{diags_ggbbbudbmn}
\end{figure}

Another example is 
\bea
\label{ggbbbudbmn}
gg \;\ra\; b u \bar{d} \bar b \mu^- \bar \nu_{\mu} b \bar b 
\eea
that is a dominant partonic process of the associated production of the top 
quark pair and higgs boson in the proton--proton collisions at the LHC,
$pp \;\ra\; t \bar t h$, where one of the top quarks decays hadronically,
the other semileptonically and the higgs boson decays into the $b\bar b$-pair,
as depicted in Figs.~\ref{diags_ggbbbudbmn}a--\ref{diags_ggbbbudbmn}c.
The diagrams depicted
in Fig.~\ref{diags_ggbbbudbmn}d--\ref{diags_ggbbbudbmn}f are 
examples of the background contributions to associated production
of the higgs boson and top quark pair. Under the same assumptions as above,
the entire number of the Feynman diagrams is equal to $67\,300$.

It is needles to say that reliable theoretical predictions for 
reactions as (\ref{ggbbudmn}), (\ref{uubbudmn}) and (\ref{ggbbbudbmn})
can be obtained only within a fully automated approach, as offered
by, e.g., {\tt carlomat} \cite{carlomat}, but it should also be possible
to solve the problem at hand with any other of
the following general purpose packages for Monte Carlo (MC) simulations:
{\tt MadGraph/MadEvent/HELAS} \cite{MADGRAPH}, 
{\tt CompHEP/CalcHEP} \cite{CompHEP}, 
{\tt ALPGEN} \cite{ALPGEN},
{\tt HELAC-PHEGAS} \cite{HELAC-PHEGAS}, 
{\tt SHERPA/Comix} \cite{SHERPA/Comix} and
{\tt O’Mega/Whizard} \cite{O’MEGA/WHIZARD}, or
the programs for calculating NLO EW or QCD corrections to scattering 
amplitudes such as
{\tt FeynArts/FormCalc} \cite{FeynForm}, 
{\tt GRACE} \cite{GRACE} and
{\tt HELAC-NLO} \cite{HELAC-NLO}.
It should be stressed that {\tt carlomat} has been 
designed first of all to deal with reactions involving heavy, unstable 
particles and it is not particularly well suited for multijet QCD processes,
mainly because of the explicit treatment of the color degrees of freedom.  
Actually, one may encounter problems with a very long computation time
already when approaching processes of 
the proton--proton scattering into 5 or more jets, e.g., computation time
of the color matrix for $q\bar q\to 5g$ with Intel Xeon 3 GHz processor 
is about 2 hours, for $gg\to 5g$ it is estimated to be about 200 hours. 
Moreover, sizes of the arrays generated in the calculations approach 
typical compiler limits.
Therefore, such processes should be much better handled with those of the 
above listed programs which use the MC summing over colors.

\section{Basic changes in the program}
In the following, the substantial changes in the program are described. 
\subsection{Phase space integration}
The MC phase space integration in {\tt carlomat} is performed with 
the use of multichannel approach which is chosen because the number of peaks
in the squared matrix element, that should be mapped out
in order to improve convergence of the integral, usually by far exceeds 
the number
of independent variables in a single parametrization of the phase space 
element given by Eqs.~(17) and (19) of \cite{carlomat}.
The peaks arise whenever the denominator of a propagator in any of the Feynman 
diagrams is approaching its minimum. In version 1 of the program, a
separate phase space parametrization is generated for each Feynman diagram
whose peaks are smoothed with appropriate mappings of the integration 
variables. A user specified number of individual phase space 
parametrizations are combined into a kinematical subroutine and then
all the subroutines are combined into a single multichannel integration
routine, as described in Sect. 2.4 of \cite{carlomat}. The number should
be chosen so as to possibly minimize the compilation time which, however,
may be quite long for multiparticle processes with large number of the Feynman
diagrams.

An improvement of the phase space integration in the current version of the
program is
based on a simple observation that the Feynman diagrams of the same topology
differ from each other only in propagators of the internal particles. This
means that the integration limits of all the variables in
phase space parametrizations corresponding to diagrams of the same topology are 
common and can be written only once.
The same holds for the Lorentz boosts of four momenta of the final state 
particles. The final state particles are divided in subsets that are 
characteristic for the topology. Their four momenta are randomly generated 
in the relative centre of mass frame of the subset and then 
boosted to the rest frame of the parent subset, and so on until they reach
the centre of mass frame of all the final state particles.

In order to illustrate this procedure let us consider the diagram of reaction
(\ref{ggbbbudbmn}) depicted in Fig.~\ref{diags_ggbbbudbmn}a.
The final state particles are divided in two subsets:
$\{\{b,\bar b\},\{b',\{u,\bar d\}\}\}$ and 
$\{\bar b',\{\mu^-,\bar{\nu}_{\mu}\}\}$, where a prime has been introduced
in order to distinguish between the identical quarks in the diagram.
The distinction makes sens, as there are 3 other dedicated phase space 
parametrizations for the diagrams that
differ from the one of Fig.~\ref{diags_ggbbbudbmn}a by the exchanges: 
$b \leftrightarrow b'$ and/or $\bar b \leftrightarrow \bar b'$.
Note, that the diagram of Fig.~\ref{diags_ggbbbudbmn}b, despite having the
same shape as that of Fig.~\ref{diags_ggbbbudbmn}a, belongs to different 
topology in a sense described in Sect.~2.1 of \cite{carlomat}.
Let us consider the subset $\{\{b,\bar b\},\{b',\{u,\bar d\}\}\}$ first.
The four momenta $p_u$ and $p_{\bar d}$ are 
randomly generated in the frame, 
where $\vec{p}_u+\vec{p}_{\bar d}=\vec{0}$. 
The four momentum $p_{b'}$ is generated in the frame, where
$\vec{p}_{b'}+\vec{p}_{\{u,\bar d\}}=\vec{0}$ and the four momenta 
$p_u$ and $p_{\bar d}$ are boosted to this frame. 
The four momenta $p_b$ and $p_{\bar b}$ are 
randomly generated in the frame, where $\vec{p}_b+\vec{p}_{\bar b}=\vec{0}$.
Then $p_b$, $p_{\bar b}$, $p_{b'}$, $p_u$ and $p_{\bar d}$ are boosted to the 
frame, where $\vec{p}_{\{b,\bar b\}}+\vec{p}_{\{b',\{u,\bar d\}\}}=\vec{0}$.
The four momenta of the second subset $\{\bar b',\{\mu^-,\bar{\nu}_{\mu}\}\}$
are generated analogously, but this time one Lorentz boost less is required.
Finally, all the four momenta of final state particles are boosted to the 
centre of mass frame. Therefore the integration limits and calls 
to the boost subroutine are written commonly for all the phase
space parametrizations corresponding to the Feynman diagrams of the same 
topology.

In the result of the modifications described above the phase space integration 
routine becomes shorter and a compilation time is reduced, by 
a factor 4--5 for multiparticle reactions,
compared to the previous version of the program.

\subsection{Hadron--hadron collisions}
Interfaces to {\tt MSTW} \cite{MSTW} and {\tt CTEQ6} \cite{CTEQ} 
parton density functions (PDFs) are added 
in the MC computation part of the program. The user should choose if she/he 
wants to calculate the cross
section of the hard scattering process at the fixed centre of mass energy, or
to fold it with the parton density functions, treating the initial state
particles as partons of either the $p\bar p$ or $pp$ scattering. This
is controlled by a single flag {\tt ihad} in the main program of the MC
computation {\tt carlocom.f}. The choice between the two supported
PDF sets is controlled with a flag {\tt ipdf}.
The program will automatically assign
integer numbers to the initial state partons according to the convention
of either {\tt MSTW} or {\tt CTEQ6}.
The user should also choose a value of {\tt iset} for the selected PDFs,
with a recommended value {\tt iset=0} (the central PDF set) for {\tt MSTW} and 
{\tt iset=3} (the leading order partons and $\alpha_s(m_Z)=0.118$) 
for {\tt CTEQ6}. Moreover, a character variable
{\tt prefix} that defines the fit order and location of the grid files must 
be specified for {\tt MSTW} PDFs.

It is also possible to choose one of 3 predefined factorization scales {\tt Q} 
by specifying a value of {\tt iscl}: \\
{\tt iscl=1 (Q=sclf*sqrt(s'))/2 (Q = sclf)/3 
(Q=sclf*sqrt(mt**2+sum of ptj**2))},\\
where {\tt sclf} is an arbitrary real (double precision) value. The user can
easily define other scales by changing the defining expressions for
{\tt qsc} in {\tt croskk.f}.

In order to avoid a mismatch of parameters, the quark masses, the minimum
value of Bjorken $x$ and a value of $\alpha_s(m_Z)$ are transferred from 
common blocks {\tt mstwCommon} of {\tt MSTW}, or
{\tt XQrange} and {\tt Masstbl} of {\tt CTEQ6} with
a single call to either {\tt mstw\_interface} or {\tt ctq6f\_interface} 
from a subroutine {\tt parfixkk} after they have been initialized  with
a call to {\tt GetAllPDFsAlt} or {\tt SetCtq6(iset)}.

If the initial state partons $p_1,p_2=0,\pm 1,\pm 2,\pm 3,\pm 4,\pm 5$
differ from each other then two contributions to the cross
section from $(p_1(x_1),p_2(x_2))$ and 
$(\mp p_2(x_1),\mp p_1(x_2))$, where ``$-(+)$'' corresponds to proton--proton
(proton--antiproton) scattering, are added in a function {\tt crosskk}.
On the other hand, a sum over different partonic initial-states that lead to the
same final state, except for the parton interchange just discussed, 
is not done automatically in {\tt carlomat}. This is because
the program generates a dedicated phase space integration routine that takes 
into account Feynman diagram topologies and peaks which
obviously may be different for some underlying partonic processes of
a considered process. Thus contributions from all the underlying partonic 
processes should be calculated separately and then they can be added, e.g.
with a program {\tt addbs} that is appended in directory {\tt test\_output}.
\subsection{Color matrix}
The color matrix in {\tt carlomat} is calculated numerically from 
the very definition, using explicit expressions for the SU(3) group structure 
constants and the Gell-Mann matrices, after its size have been reduced with the 
help of some basic SU(3) algebra properties.  In version 1.0 of 
{\tt carlomat}, the reduced color matrix was cast
in a single subroutine {\tt colsqkk.f} that was compiled together with all the 
other subroutines of the MC program and calculated anew
every time the program was run. This
posed no problem for processes with a simple color structure, but it
unnecessarily increased the computation time for processes with a big size
color matrix. 

Compilation of the large subroutine containing
the entire color matrix was very time consuming, therefore, 
in the current version of the program, a subroutine {\tt colsqkk.f} is divided
into smaller subroutines of the user controlled size which allows to compile 
much larger color matrices and speeds up the compilation process.
Moreover, computation of the color matrix is performed as a separate stage,
that is automatically executed just after the code generation. Then the 
resulting nonzero elements of the
color matrix and their labels are transferred to the directory, where they are
read from subroutine {\tt mpol2.f} of the MC program when it is executed for 
the first time.

\subsection{Cabibbo-Kobayashi-Maskawa mixing}

The CKM mixing in the quark sector is implemented in the program. 
The CKM mixing would be an unnecessary complication for many applications,
therefore, an option has been included in {\tt carlomat.f} that allows to switch
it on or off: {\tt ickm=1(yes)/else(no)}.

For the sake of simplicity, only the magnitudes of the CKM matrix elements
$V_{ij}$ \cite{PDG} are taken into account.
However, the complex phase of the CKM matrix can be easily incorporated, as 
the $W$ boson coupling to fermions that always multiplies $V_{ij}$ is complex.
To do so, it is enough
to change the type of $V_{ij}$  in {\tt inprms.f} to complex and make a 
distinction between the $Wf\bar{f}'$ couplings
and their conjugates in {\tt vertices.dat.ckm}. The latter is a new data file
that must be present in directory {\tt code\_generation} together
with {\tt vertices.dat} that defines the couplings in absence of the CKM mixing.

If the CKM mixing is included then the numbers Feynman diagrams of
reactions (\ref{ggbbudmn}) and (\ref{uubbudmn}) in the unitary gauge of 
SM, with the neglect of the higgs boson coupling to fermions
lighter than $c$-quark, increase to 596 and 1444, respectively. 
\subsection{Anomalous $Wtb$ coupling}

The top quark mass is close to the energy scale of the EW symmetry 
breaking. Therefore it is possible that the top quark coupling to the 
$W$-boson differs 
from the $V-A$ form of SM. The $Wtb$ coupling is present in any process
of the top quark production if a top quark decay in the dominant
$Wb$ channel is taken into account. Not only does it enter the 
signal Feynman diagrams of the top quark production and decay, but
it is present also in many other diagrams of the off resonance 
background contributions to the top quark production. This is illustrated
in Figs.~\ref{diags_gg} and ~\ref{diags_uu}, where the $Wtb$ coupling is
indicated by red blobs.

The effective Lagrangian of the $Wtb$ interaction containing operators of 
dimension four and five that is implemented in the current version of
the program has the following form \cite{kane}:
\bea
\label{lagr}
L_{Wtb}&=&\frac{g}{\sqrt{2}}\,V_{tb}\left[W^-_{\mu}\bar{b}\,\gamma^{\mu}
\left(f_1^L P_L +f_1^R P_R\right)t 
-\frac{1}{m_W}\partial_{\nu}W^-_{\mu}\bar{b}\,\sigma^{\mu\nu}
  \left(f_2^L P_L +f_2^R P_R\right)t\right]\nn\\
&+&\frac{g}{\sqrt{2}}\,V_{tb}^*\left[W^+_{\mu}\bar{t}\,\gamma^{\mu}
\left(\bar{f}_1^L P_L +\bar{f}_1^R P_R\right)b 
-\frac{1}{m_W}\partial_{\nu}W^+_{\mu}\bar{t}\,\sigma^{\mu\nu}
  \left(\bar{f}_2^L P_L +\bar{f}_2^R P_R\right)b\right],
\eea
where $g$ is the weak coupling constant, $V_{tb}$ is the element of
the CKM matrix, $m_W$ is the mass of the $W$ boson, 
$P_{R/L}=\frac{1}{2}(1\pm \gamma_5)$ are the chirality projectors, 
$\sigma^{\mu\nu}=\frac{i}{2}\left[\gamma^{\mu},
\gamma^{\nu}\right]$ and $f_{i}^{L}$, $f_{i}^{R}$, $\bar{f}_{i}^{L}$,
$\bar{f}_{i}^{R}$, $i=1,2$, are form factors which can be complex in general. 
The SM $Wtb$ interaction is reproduced if $f_{1}^{L}=\bar{f}_{1}^{L}=1$
and all the remaining form factors are set to 0.

The implementation of the right-handed vector coupling is straightforward,
as it is present in the neutral current interaction SM vertices
$f\bar f Z$ and $f\bar f \gamma$. However, the tensor couplings
of (\ref{lagr}) are not present in the SM and require new routines
for calculating the corresponding helicity matrix elements.
Therefore, a routine library {\tt carlolib} has been supplemented with
the following new subroutines: {\tt btwan}, {\tt btwmd}, 
{\tt bwtan}, {\tt bwtmd}, {\tt tbwan}, {\tt tbwmd},
{\tt wbtan}, {\tt wbtmd}, {\tt fefan}, {\tt ffvan}, {\tt fvfan},
{\tt vffan}. They allow to calculate all the helicity
amplitudes that are necessary for computing cross sections of any process
involving the anomalous $Wtb$ coupling defined by (\ref{lagr}) and
the lowest order top quark width $\Gamma_t$ which enters the top quark
propagator through the complex mass parameter defined in Eq.~(13) 
of \cite{carlomat}. 
If CP is conserved then the following relationships between
the form factors of (\ref{lagr}) hold:
\beq
\label{rel}
\left.\bar{f}_1^{R}\right.^*=f_1^R, 
\quad \left.\bar{f}_1^{L}\right.^*=f_1^L, \qquad \qquad
\left.\bar{f}_2^R\right.^*=f_2^L, \quad \left.\bar{f}_2^L\right.^*=f_2^R.
\eeq
$\Gamma_t$ is calculated anew, every time
the form factors $f_{i}^{L}$, $f_{i}^{R}$, $\bar{f}_{i}^{L}$ and 
$\bar{f}_{i}^{R}$, $i=1,2$, are changed. It should be stressed that for CP-odd 
choices of the form factors, i.e., if they do not satisfy (\ref{rel}), the 
widths $\Gamma_t$ of $t$ and $\Gamma_{\bar t}$ of $\bar t$ differ from each 
other. Thus, both widths are calculated and the following rule is applied
to substitute the width in the $s$-channel top quark propagator: 
$\Gamma_t$ is used if the propagator goes into $W^+b$ and 
$\Gamma_{\bar t}$ is used if the propagator goes into $W^-\bar{b}$. The rule does
not work for the propagators in $t$- or $u$-channels, but the actual value
of the top quark width should not play much of a role in them.

A new version of {\tt carlomat} allows to make predictions for the top 
quark production and decay in hadronic collisions, through different underlying 
partonic processes, while taking into account complete sets of the lowest
order Feynman diagrams and full information on spin
correlations between the top quark and its decay products. It was used
to obtain theoretical predictions in the presence of anomalous $Wtb$ coupling
for the forward-backward asymmetry 
in top quark pair production at the Tevatron \cite{afbtt} and for distributions
of $\mu^-$ in the top quark pair production reaction 
$pp \;\ra\; b u\bar{d}\;\bar{b} \mu^-\bar{\nu}_{\mu}$
at the LHC \cite{wtblhc}.
It can also be applied for studying anomalous effects in the top quark 
production and decay
in $\epm$ collisions at a linear collider \cite{ILC}, \cite{CLIC}, as
it was done before with ``hand made'' modifications of a program {\tt eett6f} 
\cite{eett6f}.

\subsection{Anomalous top--higgs Yukawa coupling}
Due to the large top quark mass, the top--higgs Yukawa coupling 
\bea
\label{gtth}
g_{t\bar th}=m_t/v, \qquad {\rm with}\qquad v=(\sqrt{2}G_F)^{-1/2}\simeq 246\;
{\rm GeV,}
\eea
is by far the biggest Yukawa coupling of SM. Its measurement, which may bring 
hints towards better understanding of the EW symmetry braking 
mechanism
of the SM, will certainly be one of the key points in the study of a profile 
of recently discovered candidate for the higgs boson \cite{higgs}.

The most general Lagrangian of $t\bar t h$ interaction including corrections
from dimension-six operators that has been implemented in the program has 
the following form \cite{aguilar}
\bea
\label{httcoupl}
\mathcal{L}_{t\bar t h}=-g_{t\bar th}\bar{t}\left(f+if'\gamma_5\right)t h,
\eea
where $f$ and $f'$ that describe the scalar and pseudoscalar departures, 
respectively, from a purely scalar top--higgs Yukawa coupling $g_{t\bar th}$ 
of SM are assumed to be real. The top--higgs Yukawa coupling of SM is 
reproduced in Eq.~(\ref{httcoupl}) for $f=1$ and $f'=0$. 

Implementation of the general coupling of Eq.~(\ref{httcoupl}) in the program
was relatively easy, as in the complex mass scheme \cite{Racoon}, the coupling 
$g_{t\bar th}$ is of the complex type.
It is parametrized in {\tt carlomat} in the following way:
\bea
\label{yuk}
g_{t\bar th}=e_W\;\frac{M_t}{2\sin\theta_W M_W},
\eea
where complex masses $M_t$ and $M_W$ are defined in Eq.~(13) of 
\cite{carlomat}, 
and the complex EW mixing parameter $\sin\theta_W$ is defined in Eq.~(14)
of \cite{carlomat}. The electric charge $e_W$ of Eq.~(\ref{yuk}) that
enters multiplicatively all the EW couplings can be 
defined as either a real or complex quantity, by selecting an appropriate
value of a flag {\tt ialpha} in {\tt carlocom.f}. 

If we now write Lagrangian (\ref{httcoupl}) in the form
\bea
\label{htt}
\mathcal{L}_{t\bar t h}=-\bar{t}\left(f_{t\bar th}^{(R)}P_R+
f_{t\bar th}^{(L)}P_R\right)t h
\eea
and take into account the form of the $\gamma_5$ and 
$P_{R/L}=\frac{1}{2}(1\pm \gamma_5)$ matrices
in the Weyl representation:
\bea
\gamma_5=\left(\begin{array}{rr}
-\mathbb{I} & 0\\
0            & \mathbb{I}\end{array}\right),\qquad
P_R=\left(\begin{array}{rr}
0 & 0\\
0 & \mathbb{I}\end{array}\right),\qquad
P_L=\left(\begin{array}{rr}
-\mathbb{I} & 0\\
0           & 0\end{array}\right),
\eea
where $\mathbb{I}$ is a $2\times 2$ unit matrix, then we will immediately see 
that
\bea
\label{fhtt}
f_{t\bar th}^{(R)}=g_{t\bar th}\left(f+if'\right), \qquad
f_{t\bar th}^{(L)}=g_{t\bar th}\left(f-if'\right).
\eea
As the SM Lagrangian of the top--higgs Yukawa interaction in {\tt carlomat} 
has the same form as Lagrangian (\ref{htt}) with
$f_{t\bar th}^{(R)}=f_{t\bar th}^{(L)}=g_{t\bar th}$, 
substitutions (\ref{fhtt}) are practically the only change that is required 
in the program.

The current version of {\tt carlomat} was used in \cite{ggtth} to study 
effects of the anomalous $t\bar t h$ coupling of Eq.~(\ref{httcoupl}) on
the differential distributions in rapidity and angles of the $\mu^-$
in the reaction $pp\;\to\; b u \bar{d} \bar b \mu^- \bar \nu_{\mu} b \bar b$
which is one of the channels of associated production of the top quark 
pair and higgs boson in proton--proton collisions at the LHC.
The dominant contribution to the reaction comes from the underlying 
gluon--gluon fusion partonic process (\ref{ggbbbudbmn}) examples of
the Feynman diagrams of which are shown in Fig.~\ref{diags_ggbbbudbmn},
where the coupling corresponding to 
Lagrangian~(\ref{httcoupl}) has been indicated with a blob.

\subsection{Scalar electrodynamics}
The knowledge of energy dependence of the total cross section of 
electron--positron 
annihilation into hadrons $\sigma_{e^+e^-\to {\rm hadrons}}(s)$ allows, through 
dispersion relations, 
for determination of hadronic contributions to the 
vacuum polarization, which in turn 
are necessary for improving precision 
of theoretical predictions for the muon anomalous magnetic moment and 
play an important role in the evolution of the fine structure constant
from the Thomson limit to high energy scales.
Theoretical predictions for $\sigma_{e^+e^-\to {\rm hadrons}}(s)$
within quantum chromodynamics are possible only in the perturbative regime
of the theory, i.e. in the high energy range, where due to the asymptotic
freedom, the strong coupling constant is small. In the energy range
below  the $J/\psi$ threshold, $\sigma_{e^+e^-\to {\rm hadrons}}(s)$ must
be measured, either by the initial beam energy scan or
with the use of a radiative return method
\cite{radret}. The idea behind the method is that the actual
energy of $e^+e^-$ scattering in the radiative reaction  
$e^+e^- \to {\rm hadrons}+\gamma\;$ 
becomes smaller if a hard photon is emitted off the initial electron 
or positron prior to their annihilation.
Thus, if the hard photon energy is measured and the photon emission off
the final state hadrons is properly modelled, it is possible to determine
the cross section of $e^+e^-\to {\rm hadrons}$ at the reduced 
energy from the corresponding radiative process
being measured at a fixed energy in the centre of mass system. At low energies,
such radiative hadronic final states consist mostly of pions, accompanied by 
one or more photons.

The scalar quantum electrodynamics (sQED) is a theoretical framework that 
allows to describe effectively the low energetic electromagnetic interaction of 
charged pions.  Despite being bound states of the electrically charged quarks,
at low energies, $\pi^{\pm}$ can be treated as pointlike particles 
and represented by a complex scalar field $\varphi$. The $U(1)$ gauge invariant
Lagrangian of sQED that is implemented in {\tt carlomat} has the following form:
\bea
\label{sqed}
\mathcal{L}_{\pi}^{\rm sQED}=
\partial_{\mu}\varphi\left(\partial^{\mu}\varphi\right)^*-m_{\pi}^2\varphi\varphi^*
-ie\left(\varphi^*\partial_{\mu}\varphi-\varphi\partial_{\mu}\varphi^*\right)
A^{\mu}
+e^2g_{\mu\nu}\varphi\varphi^*A^{\mu}A^{\nu}.
\eea
The corresponding Feynman rules can be found, e.g., in \cite{fred}. The
bound state nature of the charged pion is taken into account by
the substitutions:
$$ e\to eF_{\pi}(q^2), \qquad e^2\to e^2\left|F_{\pi}(q^2)\right|^2,$$
where $F_{\pi}(q^2)$ is the charged pion form factor \cite{fred}. 
Both three- and four-point interactions are included in the program, but
a formula for the form factor has not yet been implemented.
.

\section{Other changes and preparation for running}
Other minor changes to the program, including corrections of a few 
bugs are described in a {\tt readme} file. 

{\tt carlomat~v.~2.0} is distributed as a single {\tt tar.gz} archive
{\tt carlomat\_2.0.tgz}.
After executing a command\\
{\tt tar -xzvf carlomat\_2.0.tgz}\\
directory {\tt carlomat\_2.0} will be created.\\
For user's convenience files {\tt mstwpdf.f} of {\tt MSTW} 
and {\tt Ctq6Pdf.f} and {\tt cteq6l.tbl} of {\tt CTEQ6} are included in the 
current distribution of {\tt carlomat}. 
To put {\tt MSTW} grid files on your computer 
download a file {\tt mstw2008grids.tar.gz} from\\
{\tt http://mstwpdf.hepforge.org/code/code.html}\\
and unpack the tarball with\\
{\tt tar -xzvf mstw2008grids.tar.gz}\\
in the same directory where {\tt carlomat\_2.0.tgz} was unpacked.

In order to prepare the program for running take the following steps:
\begin{itemize}
\item select a Fortran 90 compiler in {\tt makefile}'s of {\tt code\_generation}
and {\tt mc\_computation}, which need not be the same, and compile all the 
routines stored in
{\tt carlolib} with the same compiler as that chosen in {\tt mc\_computation}; 
\item go to {\tt code\_generation}, specify the process and necessary
options in {\tt carlomat.f} and execute 
{\tt make code} from the command line. 
\end{itemize}
Once the code generation is finished,
the color matrix is compiled and computed. The control output files
{\tt test} and {\tt test\_clr} in {\tt code\_generation} should reproduce 
the delivered files {\tt test\_gg} and {\tt test\_clr\_gg} for process 
(\ref{ggbbudmn}) or {\tt test\_uu} and {\tt test\_clr\_uu} for process
(\ref{uubbudmn}). Then
\begin{itemize}
\item go to {\tt mc\_computation}, choose the required options and
the centre of mass energy by editing properly {\tt carlocom.f}
and execute {\tt make mc} in the command line.
\end{itemize}
When the run is finished all output files are automatically moved
to directory {\tt test\_output}. They should reproduce the files delivered in
directory {\tt test\_output0}. The basic output of the MC run is stored in 
a file whose name starts with {\tt tot}. Files with prefixes {\tt db} 
and {\tt dl} contain data for making plots of differential cross sections 
with boxes and lines, respectively, with the use of {\tt gnuplot}. 
The distributions of processes with different initial state partons can be 
added with the help of a simple program {\tt addbs} appended in directory
{\tt test\_output}. The program should be appropriately edited, compiled
and run twice in order to add distributions stored in files 
with prefixes {\tt db} and {\tt dl}. An example of the input file
{\tt example.plo} for {\tt gnuplot} is also appended in {\tt test\_output}.
The distributions can be plotted and viewed with a command:\\
{\tt make plot}.\\
The user can define
his/her own distributions by modifying appropriately files {\tt calcdis.f} 
and {\tt distribs.f} in directory {\tt mc\_computation}.

Note, that whenever the Fortran compiler 
in {\tt mc\_computation} is changed, or a
compiled program is transferred from a machine with different architecture,
then all the object and module files in the directory should be deleted 
by executing the commands:\\
{\tt rm *.o}\\
{\tt rm *.mod}\\
and all the Fortran files in {\tt carlolib} should be compiled anew.

Acknowledgments: This project was supported in part with financial resources 
of the Polish National Science Centre (NCN) under grant decision 
number DEC-2011/03/B/ST6/01615 and by the Research Executive 
Agency (REA) of the European Union under the Grant Agreement number 
PITN-GA-2010-264564 (LHCPhenoNet).


\begin{thebibliography}{99}
\bibitem{carlomat} K. Ko\l odziej, Comput. Phys. Commun. {\bf 180} (2009) 
                   1671;\\ 
                   K. Ko\l odziej, Acta Phys. Polon. {\bf B42} (2011) 2477.
\bibitem{MADGRAPH} T. Stelzer, W.F. Long, Comput. Phys. Commun. {\bf 81}
        (1994) 357;\\ 
F. Maltoni, T. Stelzer, 
JHEP 02 (2003) 027;\\ 
J. Alwall, M. Herquet, F. Maltoni, O. Mattelaer, T. Stelzer, 
JHEP 06 (2011) 128;\\ 
H. Murayama, I. Watanabe, K. Hagiwara, KEK-91-11.
\bibitem{CompHEP} A. Pukhov, et al., 
arXiv:hep-ph/9908288;\\
E. Boos, et al., 
Nucl. Instrum. Meth. A534 (2004) 250;\\ 
A. Belyaev, N.D. Christensen, A. Pukhov,
Comput. Phys. Commun. {\bf 184} (2013) 1729.
\bibitem{ALPGEN} M.L. Mangano, M. Moretti, F. Piccinini, R. Pittau, 
        A. Polosa, JHEP 0307 (2003) 001.
\bibitem{HELAC-PHEGAS} A. Kanaki, C.G. Papadopoulos, 
Comput. Phys. Commun. {\bf 132} (2000) 306;\\
C.G. Papadopoulos, 
Comput. Phys. Commun. {\bf 137} (2001) 247;\\
A. Cafarella, C.G. Papadopoulos, M. Worek, 
Comput. Phys. Commun. {\bf 180} (2009) 1941.
\bibitem{SHERPA/Comix} T. Gleisberg, et al., 
JHEP 0402 (2004) 056;\\
T. Gleisberg, et al., 
JHEP 0902 (2009) 007;\\
T. Gleisberg, S. H\"oche, JHEP 0812 (2008) 039. 
\bibitem{O’MEGA/WHIZARD} M. Moretti, T. Ohl, J. Reuter, 
arXiv:hep-ph/0102195;\\
W. Kilian, T. Ohl, J. Reuter, 
Eur. Phys. J. {\bf C71} (2011) 1742. 
\bibitem{FeynForm} J. K\"ublbeck, M. B\"ohm, A. Denner, 
Comput. Phys. Commun. {\bf 60} (1990) 165;\\
T. Hahn, 
Nucl. Phys. Proc. Suppl. {\bf 89} (2000) 231;\\
T. Hahn, 
Comput. Phys. Commun. {\bf 140} (2001) 418.
\bibitem{GRACE} H. Tanaka, T. Kaneko, Y. Shimizu, 
Comput. Phys. Commun. {\bf 64} (1991) 149;\\ 
F. Yuasa, et al., 
Prog. Theor. Phys. Suppl. 138 (2000) 18;\\ 
G. Belanger, et al., 
Phys. Rept. 430 (2006) 117. 
\bibitem{HELAC-NLO} G. Bevilacqua, et al., Comput. Phys. Commun. {\bf 184} 
(2013) 986.
\bibitem{MSTW}   A.D. Martin, W.J. Stirling, R. S. Thorne, G. Watt,
  "Parton distributA
  Eur. Phys. J. {\bf C63} (2009) 189-285.
\bibitem{CTEQ} J. Pumplin et al., JHEP {\bf 07} (2002) 012.
\bibitem{PDG} J. Beringer et al. (Particle Data Group), Phys. Rev. {\bf D86}
              (2012) 010001.
\bibitem{kane} G.L. Kane, G.A. Ladinsky, C.-P. Yuan, Phys. Rev. {\bf D45}
               (1992) 124.
\bibitem{afbtt} K. Ko\l odziej, Phys. Lett, {\bf B710} (2012) 671, 
                [arXiv:1110.2103].
\bibitem{wtblhc} K. Ko\l odziej, arXiv:1212.6733, accepted for publication in
Acta Phys. Pol. B.
\bibitem{ILC} James Brau, Yasuhiro Okada, Nicholas Walker, {\it et al.}
              [ILC Reference Design Report Volume 1 - Executive Summary],
              arXiv:0712.1950;\\
              J.A. Aguilar-Saavedra {\it et al.} [ECFA/DESY LC Physics
              Working Group Collaboration], arXiv:hep-ph/0106315;\\
              T.~Abe {\it et al.}, [American Linear Collider Working Group
              Collaboration],
               arXiv:hep-ex/0106056;\\
               K.~Abe {\it et al.} [ACFA Linear Collider Working Group
               Collaboration], arXiv:hep-ph/0109166.
\bibitem{CLIC} R.W. Assmann {\it et.al.} [CLIC Study Team],  CERN 2000-008;\\
               H. Braun {it et. al.} [CLIC Study Group], CERN-OPEN-2008-021, 
               CLIC-Note-764.
\bibitem{eett6f} K. Ko\l odziej, Phys. Lett. {\bf B584} (2004) 89;\\ 
                 K. Ko\l odziej, Comput. Phys. Commun. {\bf 151} (2003) 339.
\bibitem{higgs}  ATLAS Collaboration, Phys. Lett. {\bf B716} (2012) 1;\\
CMS Collaboration, Phys. Lett. {\bf B716} (2012) 30.
\bibitem{aguilar} J.A. Aguilar-Saavedra, Nucl. Phys. {\bf B821} (2009) 215
[arXiv:0904.2387].
\bibitem{Racoon} A. Denner, S. Dittmaier, M. Roth, D. Wackeroth,
                 Nucl. Phys. {B560} (1999) 33 and
                 Comput. Phys. Commun. {153} (2003) 462.
\bibitem{ggtth} K. Ko\l odziej, JHEP {\bf 07} (2013) 083, 
                [arXiv:1303.4962].
\bibitem{radret} Min-Shih Chen and P. M. Zerwas, Phys. Rev. D 11 (1975) 58.
\bibitem{fred} F. Jegerlehner, {\em The Anomalous Magnetic Moment of the Muon},
Springer, Nov. 2007, ISBN 9783540726333.
\end{thebibliography}
\end{document}